\documentclass[12pt,a4paper,final]{article}
\usepackage[pdftex]{graphicx}
\usepackage[centertags]{amsmath}
\usepackage{amssymb}
\usepackage{mathrsfs}
\usepackage{gensymb}
\usepackage{color}
\usepackage[round]{natbib}
\usepackage[a4paper]{geometry}
\usepackage{psfrag} 
\usepackage{setspace}

\newcommand{\dip}{\partial}
\newcommand{\Ren}{\text{R}}
\newcommand{\order}{\mathcal O}
\newcommand{\A}{\mathscr A}

\begin{document}

\title{\textbf{Experiments on the stability and drag of a
  flexible sheet under in-plane tension in uniform flow}}
\author{Michael T. Morris-Thomas$^{\dag}$ and Sverre Steen\\
  \small{Department of Marine Technology,
    Norwegian University of Science and Technology,}\\
  \small{Otto Nielsens vei 10, Trondheim, NO-7491, Norway.
    $^{\dag}$E-mail: \texttt{mmthomas@ntnu.no}}
}

\date{}%
\maketitle%

\begin{abstract}
  A flexible sheet in uniform parallel flow is studied in order to
  quantify its fluid dynamic drag and fluid-elastic stability
  characteristics. An experimental campaign is undertaken that
  involves a cantilevered sheet in air flow characterised by Reynolds
  numbers of order $\Ren=10^4-10^6$. The properties of the sheet
  include: constant mass per unit area; small but finite flexural
  rigidity; varying aspect ratios from within the range $0.43<l/L<1$,
  where $L$ and $l$ denote the length and width respectively; and
  tension applied at the trailing edge. The unique aspect of the
  present work is an investigation into the influence of in-plane
  tension on both the fluid drag and fluid-elastic stability of the
  sheet. In the absence of tension, the configuration resembles a flag
  and the drag coefficient is observed to decrease with increasing
  aspect ratio and Reynolds number. In the presence of tension, the
  fluid drag is significantly reduced in the region below the critical
  flow velocity at which convected wave instabilities appear. This
  critical flow velocity can be increased through the moderate
  application of in-plane tension. Under lateral tension, the drag of
  the sheet is given to good approximation by the turbulent boundary
  layer drag law for a flat plate. Once stability is lost, however,
  the drag coefficient increases rapidly with Reynolds number due to
  convected waves travelling over the sheet's surface.
  \begin{description}
  \item[Keywords:] fluid-elastic stability; fluid drag; uniform flow;
    flexible sheet; in-plane tension
  \end{description}
\end{abstract}

\onehalfspacing


\section{Introduction}

Through casual observation of a flag, otherwise known as a flexible
sheet, suspended in a breeze, the fluid-elastic instability known as
flutter is clearly visible. This problem first gained attention
through \cite{Rayleigh1879} who considered a massless tension-free
surface of infinite length in an effort to understand flag
flutter. During a flutter episode, the flag extracts energy from the
fluid, resulting in dispersive waves that originate at the leading edge
and propagate towards the trailing edge with increasing
amplitude. This oscillatory motion consequently affects the fluid
dynamic drag which is expected to increase once stability is lost and
the frontal area exposed to the flow increases. Although a flag in air
flow is a familiar example of a fluid-elastic system, there are many
other examples and applications --- the propulsion of fish
\cite[]{Daniel1984}, the dynamics of towed underwater cables
\cite[]{Dowling1988} and sheets, the biological mechanisms behind
snoring \cite[]{Huang1995}, filaments
\cite[]{Zhang2000,Schouveiler05}, paper flutter in processing and
printing \cite[]{Watanabe2002a}, drag stabilisation
\cite[]{Auman2005}, and general slender structures in axial flow
\cite[]{PaidoussisV2} to name but a few --- that motivate a study on
fluid-elastic instabilities of flexible sheets. However, what has
received limited treatment throughout available literature, is the
influence of in-plane tension on a flexible sheet in flow. Additional
rigidity through tension is expected to stabilise the sheet's fluid-elastic
response and consequently reduce its fluid dynamic drag. These effects
are investigated here.



A classical flutter-type instability of a flexible sheet of small but
finite thickness occurs when some threshold of the incident fluid
velocity or perturbation of the system is breached. The point at which
this occurs is termed the critical velocity $U_c$. In contrast to a
forced excitation, such as the reattachment of shear layers
\cite[]{Allen2001} or though mechanical means \cite[]{Taneda1974}, we
are concerned with a self-excited response. This is often
characterised by convected waves or so-called limit-cycle oscillations
\cite[see][among others]{Peskin2002,Dowell2003,Connell2007}. The
nature of the problem dictates that it is fully coupled whereby the
properties of the sheet and the surrounding fluid are equally
important and intimately linked to the sheet's response. Both
stability and the subsequent magnitude and frequency of the
instability depend on a competition between the fluid's momentum and the
sheet's inertia, and the restorative influences of the sheet's flexural
rigidity and any in-plane tension --- which can be supplied by either
the boundary layer or by external means. End-point boundary conditions
also play an important role; from free-free conditions that resemble
fish like swimming \cite[]{Farnell2005}, to clamped-free restraint
indicative of a flag suspended in flow \cite[]{FitPop01}.

To aid in our physical understanding, the character of the flow
surrounding the sheet can be divided into two separate regimes based
on the aspect ratio $\mathscr A = l/L$, where $l$ and $L$ denote the
sheet width and length respectively. For $\mathscr A \ll 1$, we
essentially have a long slender sheet and the flow about it can be
considered approximately two-dimensional in the cross-flow direction
\cite[]{Lighthill1960}. Here, flow separation at the trailing edge is
of little consequence compared to boundary layer induced tension and
added mass effects. In this regime, $U_c$ decreases as the aspect
ratio increases, and some success has been achieved at
predicting $U_c$ via a slender body formulation
\cite[see][for~instance]{LeHeLa2005,Souilliez2006}.

On the other hand, when $\A=\order(1)$, the characteristics of the
flow are far more complicated due to the three-dimensionality of the
sheet and the importance of flow separation and shear layers in the
form of an unsteady wake and downwash over the cross-flow
direction. In a strictly two-dimensional setting, the effect of an
unsteady wake on a sheet of infinite width has been considered both
theoretically \cite[see][among
others]{Huang1995,Peskin2002,ArgMah05,Tang2008} and experimentally in
the soap film experiments of \cite{Zhang2000}. The study by
\cite{Tang2008} demonstrates that the wake is of little consequence to
stability when the sheet is considered long\footnote{The term `long'
  is subjective in a two-dimensional setting. \cite{Tang2008} cite a
  mass ratio of $\rho L /M > 4$ to distinguish a long two-dimensional
  sheet --- $\rho$ being the fluid density and $M$ the sheet's mass
  per unit length.}. However, for a sheet of finite width, it is
generally understood that it will be theoretically more stable than
its two-dimensional counterpart \cite[]{Eloy2007}. The coupling of
unsteadiness with three-dimensionality has only recently been
approached numerically \cite[]{Dowell2003} with some success at
determining the critical flow velocity and the sheet's qualitative
behaviour after stability is lost.

The distinctive feature of a flexible sheet is its low flexural
rigidity. This feature complicates any experimental study where it is
important for the sheet to maintain a planar form with little or no
geometric irregularities as the flow velocity is increased up to, and
beyond, the critical velocity. To circumvent this, most authors choose
to study a sheet mounted vertically
\cite[]{Taneda1968,Datta1975,LeHeLa2005} allowing gravity to impose a
linearly varying axial tension that reduces to zero at the trailing
edge. This gravity-induced in-plane tension dominates over the minimal
amount provided by fluid friction under realistic flow conditions
\cite[]{LeHeLa2005}. Alternatively, one may mount a sheet horizontally
in flow and employ sufficient in-plane tension or so-called
pre-tension \cite[][]{Coene92,Watanabe2002a,yamaguchi2003,MMTSS2008b}
to ensure a planar sheet form up to the critical flow velocity.

Although applying in-plane tension to a flexible sheet is convenient
for experiment, it does however replicate some important practical
applications; for instance, print media and paper manufacture, and
aircraft towed advertising banners. In printing and paper manufacture,
knowledge of the stability characteristics of the sheet are important;
for towed banners on the other hand, fluid dynamic drag reduction
assumes priority over convected ripple like modes\footnote{These
  ripple like modes resemble the wrinkles which appear across the
  surface of membranes under in-plane tension
  \cite[see][]{Wong2006}.}. In the context of pure transverse tension
(cross-flow normal) applied to a simply supported sheet,
\cite{Watanabe2002a} demonstrates an elevated critical velocity by
suppressing lateral sheet displacements. Moreover, Watanabe and
colleagues present some revealing illustrations that show the flow
surrounding a sheet in the presence of pure lateral restraint to be
largely two-dimensional, distinctly unseparated, with progressive
flexural waves exhibiting a remarkable similarity to those of a
cantilevered pipe conveying fluid \cite[]{Watanabe2002a}. In contrast,
but with similar conclusions, \cite{Coene92} examined a free-free
paper sheet of $\mathscr A = \order(0.1)$ with tension supplied
axially (normal to the trailing edge). A clear increase in the
critical velocity is shown. However, given that the sheet's flexure was
not defined, coupled with the inherent anisotropy of paper, any direct
comparison with existing or new data is speculative at best.


The fluid dynamic drag of a sheet is expected to increase once
stability is lost to flutter \cite[]{Taneda1968}. This is because: the
frontal area exposed to the flow increases in the presence of flexural
waves; and the wake transforms from a well defined von K\'arm\'an
vortex street to one comprising large eddy formations
\cite[][especially Fig. 13]{Taneda1968}. While this phenomenon has
been demonstrated for laminar flows with in-plane tension provided by
gravity \cite[]{Taneda1968}, its observation in turbulent flows of
more practical interest has suffered because of geometric
irregularities of the sheet up to the point of instability
\cite[see][]{Fairthorne1930,Auman2001,Carruthers2005}. Like a flat
plate, the drag coefficient of a flexible sheet decreases with
increasing aspect ratio \cite[]{Taneda1968,Auman2001,Carruthers2005};
and, furthermore, we can expect its magnitude to approach that of a
flat plate under sufficiently small elastic deformations. To the
authors' knowledge, the effect of additional rigidity through in-plane
tension, although likely to postpone an expected rise in the drag
coefficient at the onset of a fluid-elastic instability, has hitherto
not been studied.


The present paper examines a rectangular flexible sheet of small
flexural rigidity in uniform parallel flow. The leading and trailing
edges of the sheet comprise clamped and free boundary conditions
respectively. Our primary goal is to address the issue of in-plane
tension and its impact on both the fluid-elastic stability and fluid
dynamic drag characteristics of the sheet. An experimental campaign is
undertaken whereby the sheet's dimensions are chosen such that
$0.43<\A<1$. In-plane tension is applied to the sheet at a given angle
$\theta$ to the trailing edge. The magnitude of this tension is varied
from zero (a flag) to values in the region where tension induced
rigidity is expected to dominate over flexure. The instability we are
concerned with is of the flutter-type which is characterised by
convected waves propagating from the leading to the trailing edge of
the sheet's surface. However, under certain conditions transverse mode
shapes are possible. The paper is organised as follows: appropriate
scalings and an approximate analytical model describing the
instability of a flexible sheet are presented in \S~\ref{sec:approx};
the experimental campaign undertaken is described in
\S~\ref{sec:exps}; results pertaining to the flutter-type instability
\S~\ref{sec:UC}, the fluid dynamic drag \S~\ref{sec:FD} and the fluid
dynamic drag in the presence of in-plane tension \S~\ref{sec:tension}
are then discussed; and finally, conclusions are provided in
\S~\ref{sec:conc}.

\section{Scaling and a qualitative model}\label{sec:approx}

We first consider a simple two-dimensional analytical model of a
flexible sheet. This enables us to construct a consistent set of
dimensionless parameters, and facilitates an understanding of how
these parameters affect a flutter-type instability. This approach is
motivated by \cite{Coene92} and \cite{SVZ05} and we combine their
ideas to include linear elastic deformation of the sheet due to both
flexure and axial tension.

We consider a flexible sheet of length $L$ comprising an isotropic
material of mass per unit area $M$ immersed in an ideal fluid in the
presence of uniform parallel flow of velocity $U$. The flow approaches
from $X=-\infty$ and we assume that the sheet occupies the plane $Y=0$
in its quiescent position. The vertical displacement of the sheet from
its mean position is denoted by $w(X,t)$, which we assume small --- $w
\ll \lambda$, $\lambda$ being a typical wavelength, or $w \ll L$, $L$
being the sheet length. This assumption is justified given that we are
only interested in the stability criteria, rather than post-stability
behaviour, of the sheet. Under the restriction of linear elasticity,
the vertical displacement of the flexible sheet is governed by
\begin{equation}
\label{eq:age}
   B \frac{\dip^4 w}{ \dip X^4} - T \frac{\dip^2 w}{\dip X^2} + M
   \frac{\dip^2 w}{\dip^2 t} = \Delta P(X;t), 
\end{equation}
which is essentially the classical Euler-Bernoulli beam equation with
an additional term accounting for axial tension per unit width
of the sheet $T$. The fluid pressure acting across the sheet is
denoted $\Delta P(X;t)$ and $B=Eh^3/12(1-\nu^2)$ defines the
flexural rigidity, where $h$ is the sheet thickness, $E$ the elastic
modulus, and $\nu$ denotes Poisson's ratio.

We now assume a harmonic response of the sheet, and that the amplitude
of this response is much smaller than the sheet length. Its
linear displacement from the mean position $Y=0$ can then be written
\begin{equation}
  \label{eq:eta}
  w(X,t) = A \exp(iKX - i\Omega t),
\end{equation}
where $K$ denotes the spatial wavenumber and $\Omega$ the circular
wave frequency. In adopting (\ref{eq:eta}), we are essentially
implying that $\lambda \ll L$, which is an obvious
simplification. However, this is justified given our motivation
involves identifying consistent scaling parameters and deducing their
qualitative contribution to the sheet's stability.

Under the assumption that the sheet motion and zero flux boundary
condition are linear about the plane $Y=0$, the appropriate form of
the fluid pressure, in accordance with (\ref{eq:eta}), acting across
the sheet is \cite[cf.][Art.~232]{Lamb1932}
\begin{equation}
  \label{eq:Delp}
  \Delta P(X;t) = 2 \frac{\rho}{|K|} (\Omega - UK)^2\, w,
\end{equation}
where $|K| \equiv \sqrt{K^2}$. In scaling the governing equation, it is
convenient to adopt:
\begin{equation}
  X = x L, \quad w = \eta L, \quad t = \tau \frac{L}{U}, \quad K =
  \frac{k}{L}, \quad 
  \Omega = \omega \frac{U}{L}, \quad \Delta P =  \rho U^2 \Delta p.    
\end{equation}
Introducing these into the governing equation (\ref{eq:age}) with
(\ref{eq:Delp}) provides 
\begin{equation}
  \label{eq:gescaled}
  \eta_{\tau \tau}
  + \frac{1}{v^2} \eta_{xxxx}
  - \kappa \eta_{xx}
  = 2 \frac{\mu}{|k|} (\omega - k)^2 \, \eta,
\end{equation}
which describes the displacement of the sheet in terms of three
dimensionless quantities defined as follows: 
\begin{equation}
\label{eq:dqs}
  \mu = \frac{\rho L}{M}, \qquad 
  v = U \left(\frac{M L^2}{B} \right)^{1/2}, \qquad 
  \kappa = \frac{T}{M U^2}.
\end{equation}
The first parameter, $\mu$, describes the ratio of fluid forcing to
the inertia of the sheet with time scaled as $t(U/L)$, and the second,
$v$, is commonly referred to as the reduced velocity and is
essentially the ratio of time scales between the elastic deformation
due to flexure and that which results from the fluid's forcing
\cite[]{Tang2008}. The third parameter, $\kappa$, is essentially the
ratio of time scales between the elastic deformation due to fluid
forcing and the elastic deformation due to tension. These three
parameters can be combined to produce:
\begin{equation}
  \label{eq:vbeta}
  \frac{v}{\mu} = U \biggl(\frac{M^3}{\rho^2 B}\biggr)^{1/2}, \qquad
  \beta = \kappa v^2 = \frac{T L^2}{B}. 
\end{equation}
The utility of which is that $v/\mu$ effectively removes the
dependence on sheet length from the scaled fluid velocity. This form
was employed by \cite{Tang2008} and it is also the appropriate
velocity scaling for an infinitely long flexible plate under fluid
loading \cite[cf.][]{CriOsw91}. The advantage of $\beta$ is that we
have effectively removed the dependence on fluid velocity from the
dimensionless tension so that we now have a direct measure of the
deformation due to tension with that from flexure. This form appears
in the study conducted by \cite{yamaguchi2003} for simply supported
sheets. We should also point out that although $\beta$ in (\ref{eq:vbeta})
is derived by assuming that $T$ acts in the axial direction, we can
also employ this definition for $\beta$ when $T$ is applied at an
angle $\theta$ to the trailing edge of a three-dimensional sheet
(cf. \S~\ref{sec:exps}). In which case, $T$ is understood to be the
magnitude of the in-plane tension.

We now construct a dispersion relation connecting the normalised
circular frequency $\omega$ and wave number $k$ by placing
(\ref{eq:eta}) in (\ref{eq:gescaled}), whence 
\begin{equation}
  \label{eq:D}
  \mathscr{D}(k,\omega) = \omega^{2} -
  \frac{1}{ v^{2} } ( {k}^{4} + \beta\,{k}^{2} )
  + 2\,\frac {\mu}{ \left| k \right| }\, \left( \omega-k \right) ^{2}
  = 0,  
\end{equation}
where $\beta = \kappa v^2$. A dispersion relation of similar meaning to
(\ref{eq:D}) is presented in \cite{SVZ05} with axial tension omitted
but intimated in the form of a Blasius friction law. By replacing
$\beta$ with $\beta=1.3 \Ren^{-1/2} L^2/B$, where $\Ren$ denotes the
Reynolds number, this later form appears in \cite{Connell2007} for
constant $\Ren$. Here, however, we leave the magnitude of $\beta$
unspecified.

In solving the dispersion relation for $\omega$ we arrive at
\begin{equation}
  \label{eq:omega}
  \omega^{\pm} = 
  \frac {k}{ \left| k \right| + 2\,\mu} 
  \biggl\{ 2 \mu \pm \frac { \sqrt { \left|k \right| } }{v} 
 \biggl[ {k}^{2}
  \left| k \right| +\beta\, \left| k \right| - 
  2\,\mu\,{v}^{2}+2\,\mu\,{k}^{2}+2\,\mu\,\beta \biggr]^{1/2} \biggr\}
\end{equation}
for any real-valued $k=n\pi$ with $n\in \mathbb{N}$. It is interesting
to note that (\ref{eq:omega}) is not too dissimilar to the dispersion
relation for a Kelvin-Helmholtz type instability of a two layer
inviscid fluid, whereby the restorative influences of sheet flexure
and axial tension are analogous to buoyancy and surface tension
respectively.

To determine instability, we search for frequencies which exhibit
spatially growing modes whereby $\Im (\omega)>0$ is required. Consequently,
in (\ref{eq:omega}) we set the terms enclosed in square braces to zero
and find that
\begin{equation}
  \label{eq:v_c}
v_c^2 > \left(  1 + \left| k \right| / 2 \mu  \right) 
 \left( {k}^{2}+\beta \right)
\end{equation}
is required to realise instability of the flutter-type through
convected flexural waves. 

\begin{figure}[t]
  \centering
  \includegraphics[scale=1]{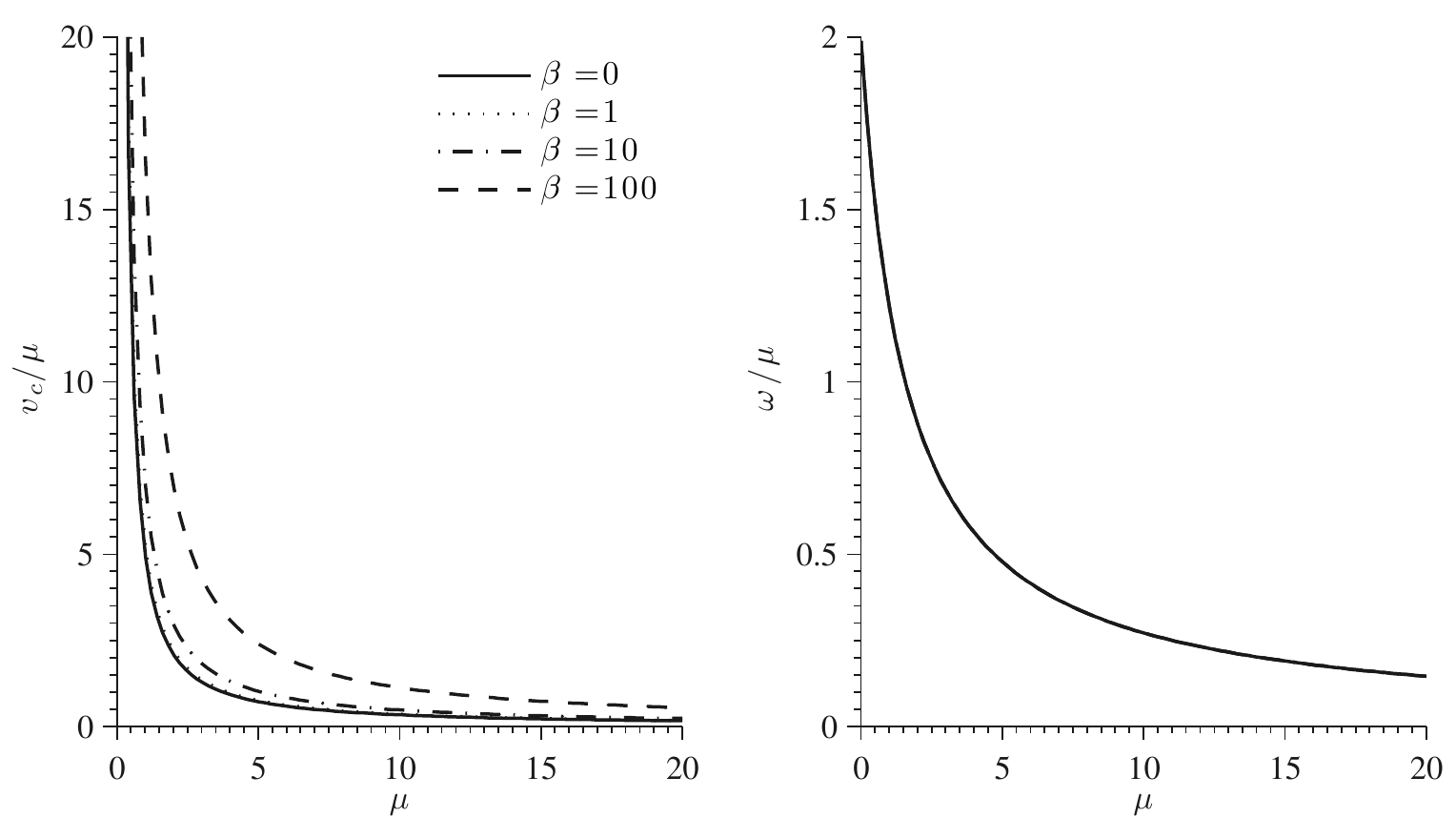}
  \caption{Left axis: the critical velocity $v_c/\mu= U (M^3 / \rho^2
    B)^{1/2}$ plotted against the mass ratio $\mu = \rho L/M$ for the
    first eigenmode $k=\pi$ of a flexible sheet under axial
    tension $\beta=\kappa v^2$ in uniform flow. Right axis: the
    corresponding circular frequency $\omega/\mu = \Omega ( M / \rho
    U)$ at the point of instability.}
  \label{fig:disp}
\end{figure}

Equations (\ref{eq:omega}) and (\ref{eq:v_c}) with $k=\pi$ are plotted
in Figure \ref{fig:disp}, where we have chosen to scale the ordinate of
each by $\mu$ which conveniently collapses the data. The role of
$\beta$ is to postpone instability; so that for $\beta=100$, for
instance, $v_c/\mu$ has increased by a factor of approximately 3.5
over $\beta=0$. The fact that $\beta$ does not affect $\omega/\mu$ at
the point of instability will come as no surprise, given the form of
(\ref{eq:omega}).

For practical purposes we can write (\ref{eq:v_c}) in dimensional
form,
\begin{equation}
  \label{eq:U_c}
  U_c^2 > \frac{|K|}{2 \rho M } (M  + 2\rho/ |K|)( B K^2 + T),
\end{equation}  
which serves to illustrate the effects of both structural and added
mass, flexure, and axial tension on stability.  

Despite their simplicity, equations (\ref{eq:omega}), (\ref{eq:v_c})
and (\ref{eq:U_c}) illustrate some important aspects of the
system. Firstly, any increase in the length of the sheet will
destabilise the system --- this effect is of $\order(L^3)$ which is
quite disastrous from a practical standpoint. Both flexural rigidity
and tension act to stabilise the sheet and any increase will be
coupled by an increase in both $U_c$ and $\omega$ meaning the fluid
must supply more energy to excite a fluid-elastic response. On the
other hand, an increase in either structural or fluid added mass,
determined by $2\rho/ |K|$, will promote instability. Another
important point, first mentioned by \cite{Peskin2002} and more
recently by \cite{Connell2007}, is that a massless system is
infinitely stable. This is clearly demonstrated by (\ref{eq:U_c})
where $U_c \rightarrow \infty$ as $M \rightarrow 0$. Lastly, it is
worth pointing out that with axial tension replaced by a constant
flow-induced tension from a Blasius boundary layer drag law
\cite[cf.][]{Connell2007}, it contributes $\beta \approx \order(1)$
and therefore provides minimal additional stability when compared to
an externally applied tension.

\section{Experiments}\label{sec:exps}

\begin{figure}[b]
  \centering
  \includegraphics{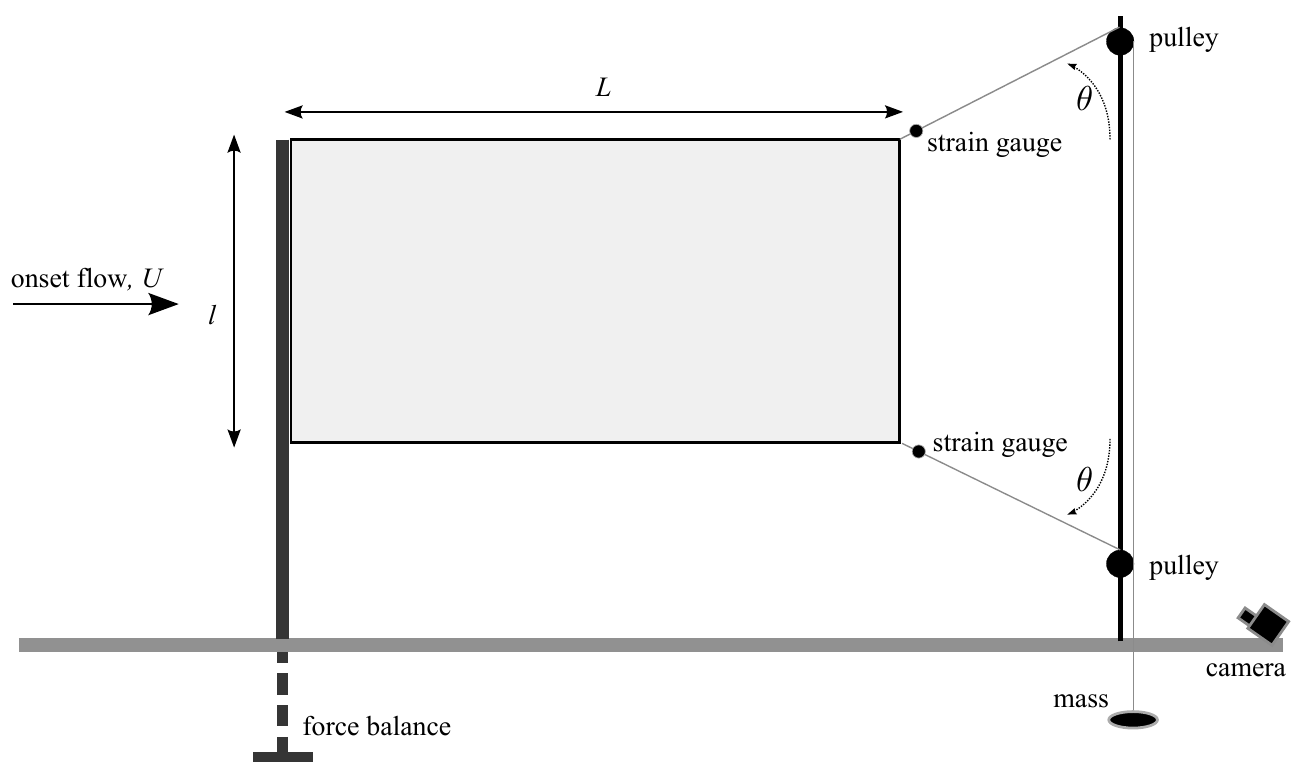}
  \caption{A schematic of the test set-up. The wind tunnel floor is
    represented by the grey horizontal line.}
  \label{fig:diagram}
\end{figure}

We now outline the experimental campaign undertaken to investigate the
drag and stability of a rectangular flexible sheet in uniform
flow. The experimental campaign was conducted in the low speed wind
tunnel of the Aerodynamics Laboratory at the Norwegian University of
Science and Technology. The tunnel comprises an available test-section
of 2.7m $\times$ 1.83m and allows fluid velocities up to 30ms$^{-1}$
with a turbulence intensity of less than 5\%. The sheet was mounted
vertically along the centre-line of the test-section with a clamped
leading edge and free trailing edge. A diagram of the test set-up is
provided in Figure \ref{fig:diagram} which we shall now explain in the
following subsections.

\subsection{Apparatus}

A transparent Polyethylene material of thickness $h=0.15$mm was chosen
for the sheet as it provided a reasonably low mass per unit area
$m=0.141$kgm$^{-2}$ and sufficient bending rigidity $B=59.65 \times
10^{-6}$Nm$^2$/m to allow an observation of flexural waves. Using a
width of $l=75$cm, the length of the sheet was varied to produce five
aspect ratios from within the range $0.43<\A<1$. These five geometries
are hereafter denoted S1, S2, S3, S4 and S5. The characteristics of
which are summarised in Table \ref{tab:parameters}. With these
dimensions, Reynolds numbers were in the order of $\Ren = 10^4-10^6$
--- a range which encompasses the transitional regime for the boundary
layer on a flat plate. In addition to the aspect ratio, the other
parameters varied were the magnitude of the in-plane tension $T$ and
the angle $\theta$ at which it was applied to the trailing edge.

At the corners of the trailing edge, tension was applied. This was
accomplished with two lengths of Kevlar string attached to two sets of
known and equal weights. These weights were parametrically varied to
produce a combined pre-tension in the order of $1-15$N. To achieve a
constant mean tension throughout a test run, each string travelled
through a pulley system that directed it underneath the wind tunnel
where it was attached to the weights. The position of the pulleys was
varied to produce three different angles at which the string left the
trailing edge: $\theta = 0$, $22.5\degree$, and $45\degree$; whereby
$\theta=0$ implies a perpendicular tension to the trailing edge. The
tension in each string was measured directly behind the trailing edge
of the sheet with two ring type strain gauges surrounded by swivels to
avoid excessive rotations. The strain gauges were calibrated with an
expected error in the order of $\pm0.01$N. The applied tension produced
a small oscillatory out-of-plane component when the sheet experienced
flutter. At worst, its amplitude was around 7\% of the
total in-plane tension at $U=12$ms$^{-1}$ for S5. The magnitude of the
out-of-plane component decreased for increasing $\A$ such that for S1
and S2 its amplitude was less than $0.01T$ at the maximum flow velocity
considered here.

A three-component force balance, situated underneath the centre of the
test-section, was employed to measure the in-line force. This was
accomplished by utilising a vertical pole, cantilevered from the force
balance, that extended into the centre of the test-section. Along the
leading edge, nylon tape was employed, affixing the sheet to the pole
and simulating a clamped boundary condition. To minimise its drag
contribution and affect on the fluid-elastic deformation of the sheet,
the vertical pole, of height 1.43m, comprised an elliptical
cross-section ($a=4$cm and $b=2$cm). The bottom of each sheet was
located 54cm above the base of the wind tunnel. This was deemed
sufficient to avoid the developing turbulent boundary layer along the
wind tunnel floor and walls --- of thickness $\delta \approx 15$cm at
the leading edge location, and, assuming a logarithmic velocity
profile, climbing to approximately 17cm at the trailing edge position.

The incident flow velocity was obtained via a Pitot-static tube fixed
28cm from the roof of the wind tunnel. At an air temperature of
22.1\degree C, the properties of the fluid used throughout the
experiments are $\rho=1.205$kgm$^{-3}$ and
$\nu=0.150$cm$^2$s$^{-1}$. Measurements from the Pitot-static tube,
force balance and strain gauge channels were acquired through a HBM
instrumentation amplifier attached to a notebook computer running
customised data acquisition software. All data was digitised at 200Hz
and pre-processed through a low-pass Butterworth filter at 40Hz to
avoid ambient noise. In addition, each run was recorded at 30 frames
per second with a digital camera situated 44cm behind the trailing
edge position on the wind tunnel floor.

Initial testing revealed the drag coefficient of the mounting pole to
be $C_D=0.59$. Hence, the independent drag contribution of the pole
ranged from 0.003N at 0.5ms$^{-1}$ quadratically climbing to a maximum
of 1.59N at 12ms$^{-1}$ --- the largest fluid velocity examined
here. We expect that the addition of the attached sheet to the pole
has the effect of a splitter plate and will therefore reduce the drag
contribution of the pole further. Using a circular cylinder and
splitter plate configuration as a guide \cite[\S~3-6]{Hoerner1965}, we
can infer a conservative lower bound of $0.6C_D$ for the portion of
the pole comprising the attached sheet. We can therefore expect a drag
reduction by the presence of the sheet by, at most, 21\% of the total
pole drag. To place this into perspective, this amounts to, at worst,
approximately 3\% of the total measured drag force of the sheet at
$U=12$ms$^{-1}$. Moreover, we suspect that given the elliptical
cross-section of the pole, combined with the flexibility of the sheet
allowing some momentum exchange in the wake at the onset of flutter,
the drag reduction experienced by the pole will be somewhat smaller
than our 21\% upper bound. This effect is not corrected for in the
data analysis implying that the calculated drag of the sheet will be
slightly smaller than expected.
 
A typical test run commenced from rest and the flow velocity of the
wind tunnel gradually increased over increments of approximately
0.2-0.5ms$^{-1}$. This continued beyond the onset of the first
observed convected wave mode, until wild oscillations of the trailing
edge ensued. At that point, the flow velocity was either terminated or
gradually decreased over increments to cover the fluid-elastic
instability region. Each run lasted for approximately 12-20 minutes,
with time windows for each chosen fluid velocity, on average, 42.1s in
length with a standard deviation of 10.6s. This process was repeated
over the parameter space of the test matrix which included five sheet
geometries, three angles, and four to seven different magnitudes of
trailing edge tension.

\subsection{Data analysis and processing}

Each data set was separated into time windows of length determined
from the velocity profile incident on the sheet. After windowing, a
general Fourier transform was performed to determine the average flow
velocity, in-line force, and trailing edge tension for each time
window. Moreover, the available frequency information allowed us to
determine the dominant harmonics of the trailing edge tension and
in-line force. The fluid-elastic instability was clearly observable
from the trailing edge tension where the magnitude of the first harmonic
increased by a factor of approximately 10 over one velocity increment
once stability was lost. At this point, the Strouhal number, based on
the length of the sheet ($\mathrm{St}=fL/U$), was in the order of
$0.4-0.6$ --- around half the value reported by \citet{SVZ05} for a
cantilevered sheet in water with $\beta=0$. As the flow velocity was
increased beyond critical, $\mathrm{St}$ increased in a linear fashion
to a peak of around $2.2-3.0$ with the lower limit occupied by data
sets comprising the largest in-plane tensions explored here.

The drag experienced by the sheet was determined from the in-line
force measured at the force balance with the drag from the pole and
total horizontal tension at the trailing edge subtracted. The drag
force, denoted $F_D$, was then normalised according to $C_D = F_D /
0.5 \rho A U^2$ where $A$ is the surface area of the sheet and $U$ the
flow velocity. The flow velocity, determined from the pressure
measured by a Pitot-static tube, over each time window comprised a
standard deviation of less than 0.05ms$^{-1}$ from the mean for
$U>2.4$ ms$^{-1}$. Below this range, however, the flow velocity showed
some variability with the standard deviation increasing to
0.19ms$^{-1}$ in some cases.

\section{Results and discussion}\label{sec:results}

\begin{figure}[p]
  \centering
  \includegraphics[width=\linewidth]{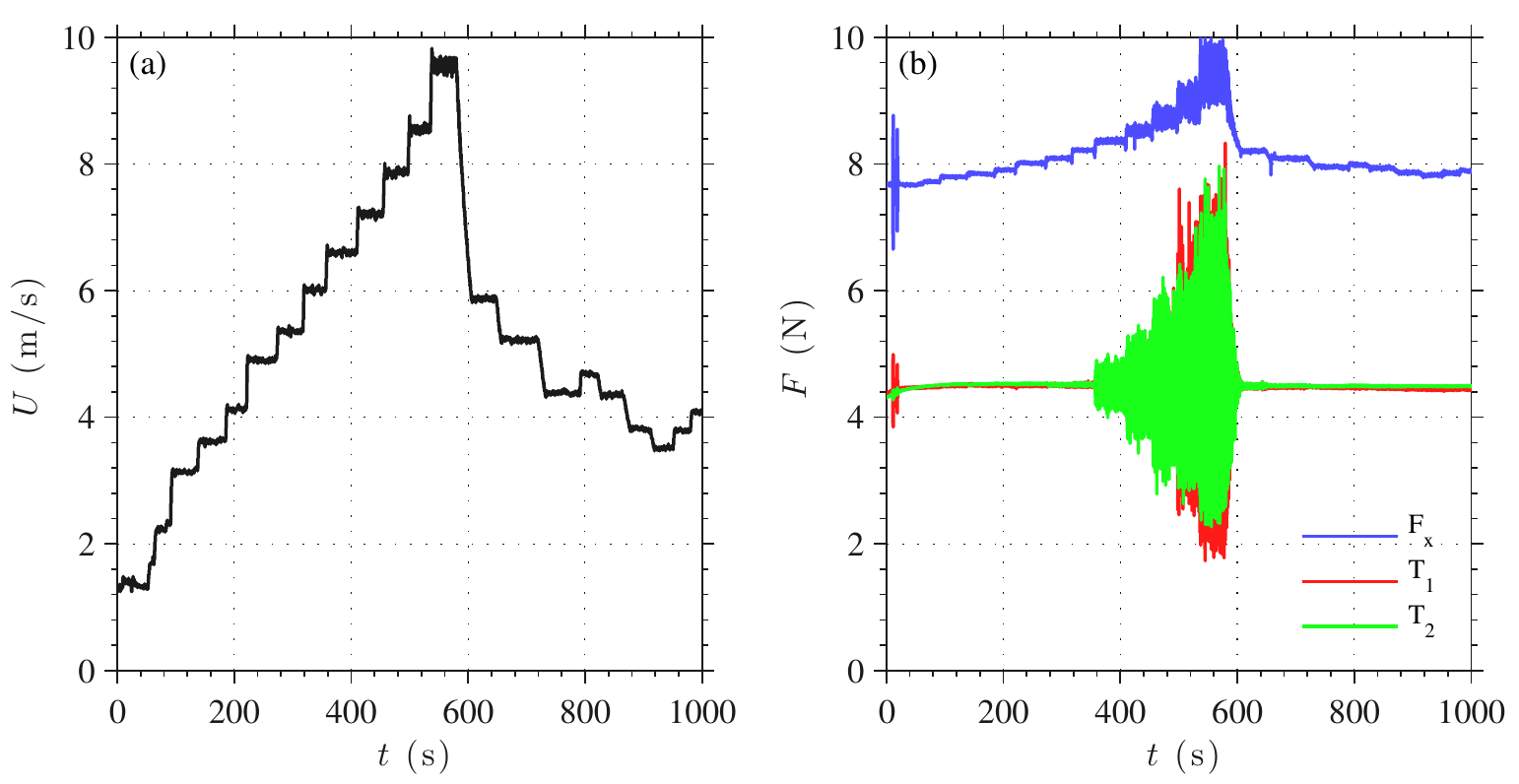} 
  \caption{A typical time series for S2 (0.75m$\times$1.25m) with
    tension applied at 22.5\degree to the trailing edge. (a)
    illustrates the incident flow velocity and (b) shows the forces
    measured: in-line $\mathrm{F_x}$; at the upper strain gauge
    $\mathrm{T_1}$; and at the lower strain gauge $\mathrm{T_2}$.}
  \label{fig:ts}
\end{figure}

A selected run from the experiments is illustrated in Figure
\ref{fig:ts} which shows the incremental increase in fluid velocity
and subsequent oscillatory behaviour of the forces measured in-line
and at the trailing edge once stability is lost to flutter. The
behaviour illustrated in the plot once stability is lost, matches our
observation of surface waves convected from the leading to the
trailing edge of the sheet. The time series results presented in
Figure \ref{fig:ts} are quite typical of all cases considered
throughout the experimental campaign. Frequency analysis of the
time-windowed data indicates that the first modal frequency of the
dominant convected wave event is in the order of 1-10Hz at the point
of instability.  The effect of which is transferred to both the force
balance and strain gauges at the trailing edge (cf. Figure
\ref{fig:ts}). Rather than growing in time, these oscillations, which
are steady and well defined, grow in amplitude and frequency as the
fluid velocity is increased further beyond the first observed
instability. When this occurs, a linear dispersion relation can no
longer relate the frequency and wavenumber of the flexural waves as
elastic nonlinearities, such as curvature, become important. These
nonlinearities will act to limit the amplitude of the instability. One
could argue that this behaviour is in fact another stable state
occupied before the sheet undergoes chaotic motion for flow velocities
much larger than critical. However, we are not concerned with this
chaotic motion at large fluid velocities here.

\subsection{Fluid-elastic instability}\label{sec:UC}

\begin{figure}[p]
  \centering
  \includegraphics{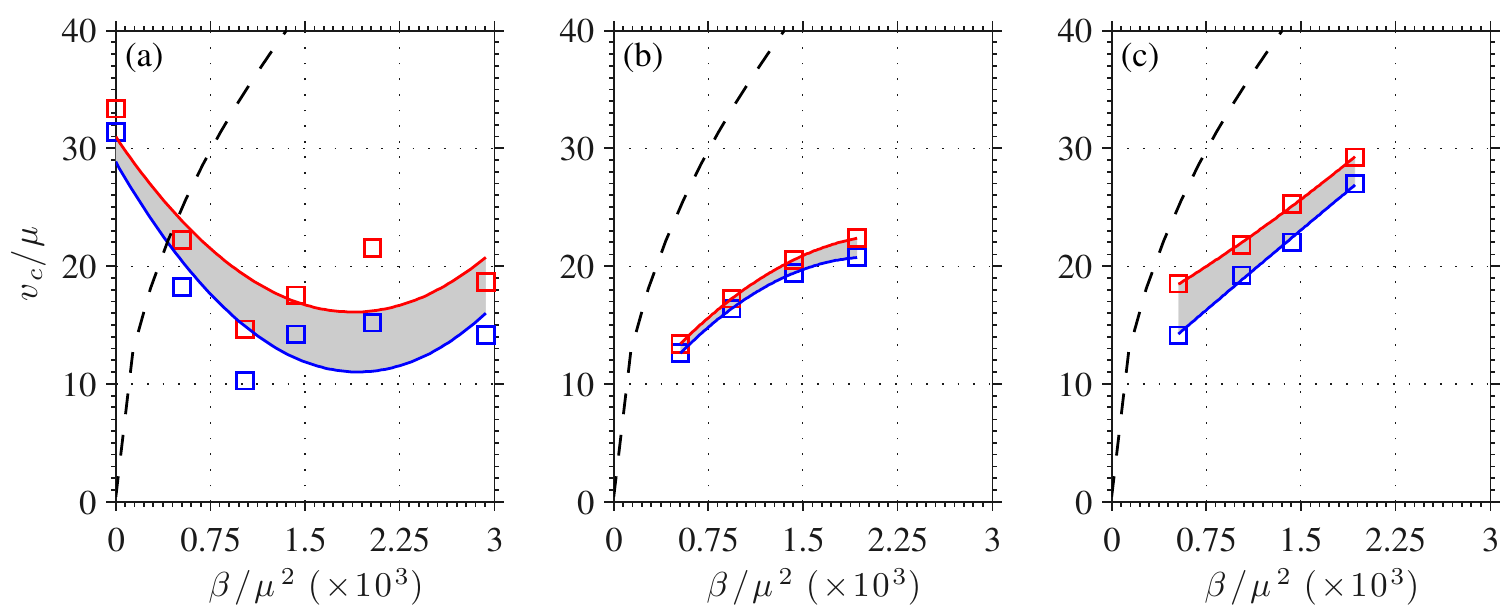}
  \caption{S2: $\A=0.75$. The measured critical flow velocity
    $v_c/\mu= U (M^3 / \rho^2 B)^{1/2}$ plotted against trailing edge
    tension $\beta / \mu^2 = T ( M^2 /\rho^2 B)$ applied at angles of:
    (a) $\theta=0$; (b) $\theta=22.5\degree$; and (c)
    $\theta=45\degree$. The dashed line corresponds to
    Eq. (\ref{eq:v_c}).}
  \label{fig:S2_vc}
\end{figure}

\begin{figure}[p]
  \centering
  \includegraphics{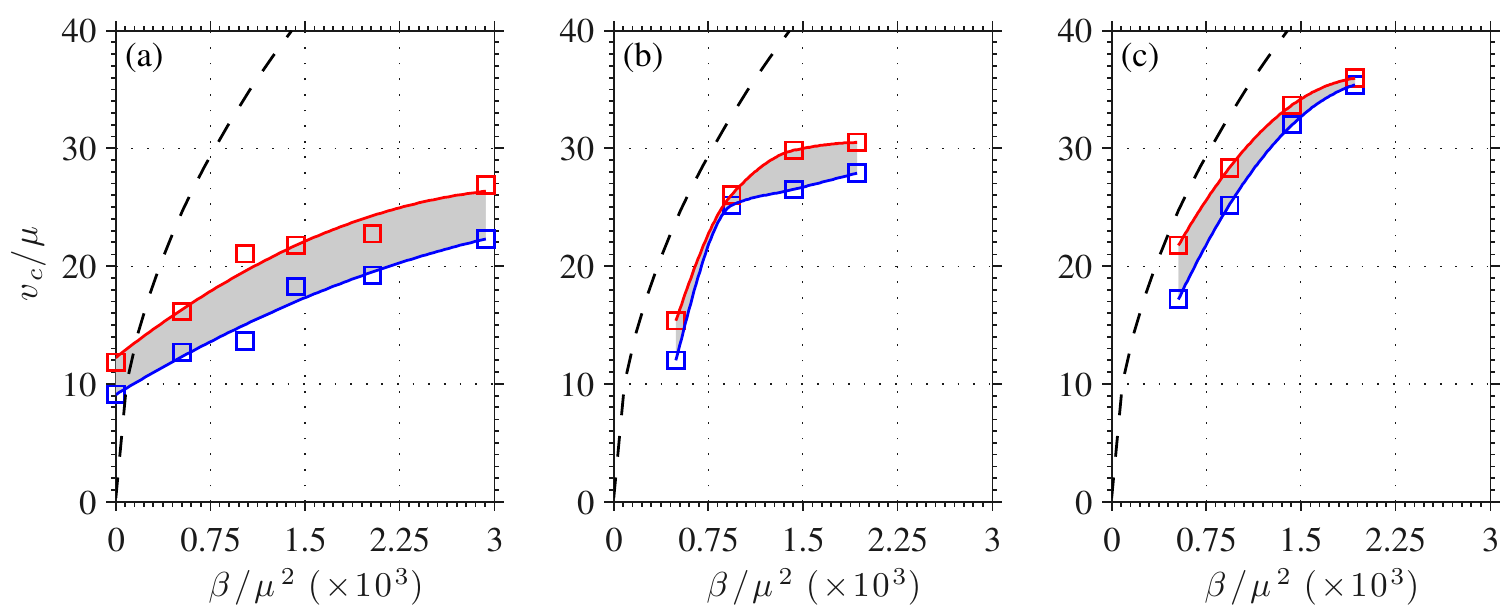}
  \caption{S3: $\A=0.60$. The measured critical flow velocity
    $v_c/\mu= U (M^3 / \rho^2 B)^{1/2}$ plotted against trailing edge
    tension $\beta / \mu^2 = T ( M^2 /\rho^2 B)$ applied at angles of:
    (a) $\theta=0$; (b) $\theta=22.5\degree$; and (c)
    $\theta=45\degree$. The dashed line corresponds to
    Eq. (\ref{eq:v_c}).}
  \label{fig:S3_vc}
\end{figure}

\begin{figure}[p]
  \centering
  \includegraphics{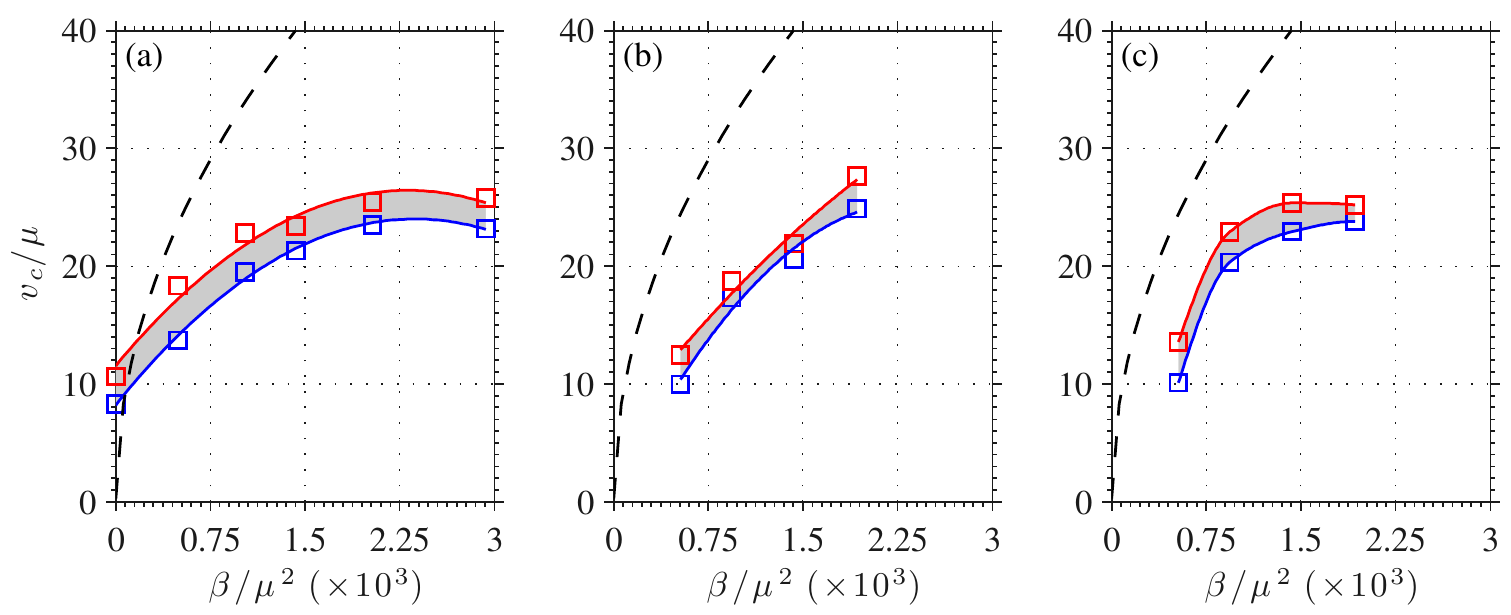}
  \caption{S4: $\A=0.50$. The measured critical flow velocity
    $v_c/\mu= U (M^3 / \rho^2 B)^{1/2}$ plotted against trailing edge
    tension $\beta / \mu^2 = T ( M^2 /\rho^2 B)$ applied at angles of:
    (a) $\theta=0$; (b) $\theta=22.5\degree$; and (c)
    $\theta=45\degree$. The dashed line corresponds to
    Eq. (\ref{eq:v_c}).}
  \label{fig:S4_vc}
\end{figure}

We first examine loss of stability due to flutter. The critical
velocity, denoted by $v_c$, was determined by examining the
first-harmonic of the trailing edge tension. A sharp jump in magnitude
and phase shift of the first harmonic indicates stability loss and the
appearance of convected waves. Figures \ref{fig:S2_vc},
\ref{fig:S3_vc} and \ref{fig:S4_vc} illustrate the results obtained
from this analysis for S2, S3 and S4 respectively. In each figure, the
scaled velocity $v_c/\mu$ is plotted against the trailing edge tension
for $\theta = 0$, $22.5\degree$, and $45\degree$. We represent the
tension by the parameter combination $\beta / \mu^2 = T ( M^2 /\rho^2
B)$, where $T$ is the tension applied by the weight and the
multiplicative factor $M^2/\rho^2 B$ is constant throughout; the
utility of which removes the dependence of sheet length and fluid
velocity from the scaled tension and allows us to examine the effect
of pure tension across each sheet considered. For comparative
purposes, we include the stability boundary determined from the linear
dispersion relation (\ref{eq:v_c}) for the first oscillatory mode
$k=\pi$.

It is important to note that the exact instability boundary is not
definitive in practice. There is a small but finite range of fluid
velocities at which the sheet can undergo periods of both stable and
unstable oscillatory motion. The transferal between these two states
does not appear to be related to any of the input variables examined
here. This bi-stability could be memory-dependent, such as that
demonstrated numerically by \cite{Peskin2002} and experimentally by
both \cite{Watanabe2002a} and \cite{Souilliez2006}, or perhaps it may
originate from perturbations of the flow velocity and any residual
turbulence available. The dynamics of the system within this region are
unclear and, in Figures \ref{fig:S2_vc}, \ref{fig:S3_vc} and
\ref{fig:S4_vc}, we account for this uncertainty by incorporating a
shaded region where upper and lower bounds of $v_c$ are indicated by
shape preserving polynomial interpolants.

The experimental results for $v_c$ indicate that instability can be
postponed by increasing the in-plane tension. In most cases we observe
a monotonic increase in $v_c/\mu$ with increasing $\beta /
\mu^2$. Apart from a persistent vertical offset, this trend is
partially captured by the qualitative expression for $v_c$
(\ref{eq:v_c}). We suspect that this offset emerges from our limited
description of the fluid added mass which appears in the dispersion
relation. The qualitative theory assumes that the fluid pressure is
composed of one single spatial mode that obeys only a periodicity
condition at the sheet end-points. In reality, however, the the first
modal frequency of the instability will comprise a number of spatial
mode shapes.

In contrast to the above, the results pertaining to S2 with tension
applied at $\theta=0$ (cf. Figure \ref{fig:S2_vc}a) follow an entirely
different trend. In particular, $v_c$ measured at values of $\beta /
\mu^2 = 0 $ and $400$ are exceptionally large in comparison to their
companion data. The dynamics of the instability appear to be different
and we present Figure \ref{fig:S2images} to illustrate snap-shots of
S2 with $\beta / \mu^2 = 400$ both before and after the appearance
of convected waves. Although the left frame of Figure
\ref{fig:S2images} shows S2 to be stable to convected waves, we
observe a transverse oscillation mode. The right frame, at a higher
fluid velocity, clearly shows this transverse standing wave envelope
which is now superimposed on a convected wave instability. Its wave
length is approximately $l/2$. These well defined transverse modes
were observed only for the shortest sheets, $\A= 0.75$ and 1, where
static deflection of the sheet due to its weight is minimal. Not
surprisingly, with in-plane tension applied laterally
($\theta=22.5\degree$ and $45\degree$), these transverse modes are
suppressed (cf. Figures \ref{fig:S2_vc}b and
\ref{fig:S2_vc}c). \cite{Souilliez2006}, in the context of flags with
$\beta=0$, suggests that such transverse modes and
three-dimensionality can delay the onset of flutter by artificially
increasing the rigidity of the system. This may well be the case
because out-of-plane bending will increase the stiffness of the sheet
to in-plane bending or, in our case, instability.

\begin{figure}[p]
  \centering
  \includegraphics{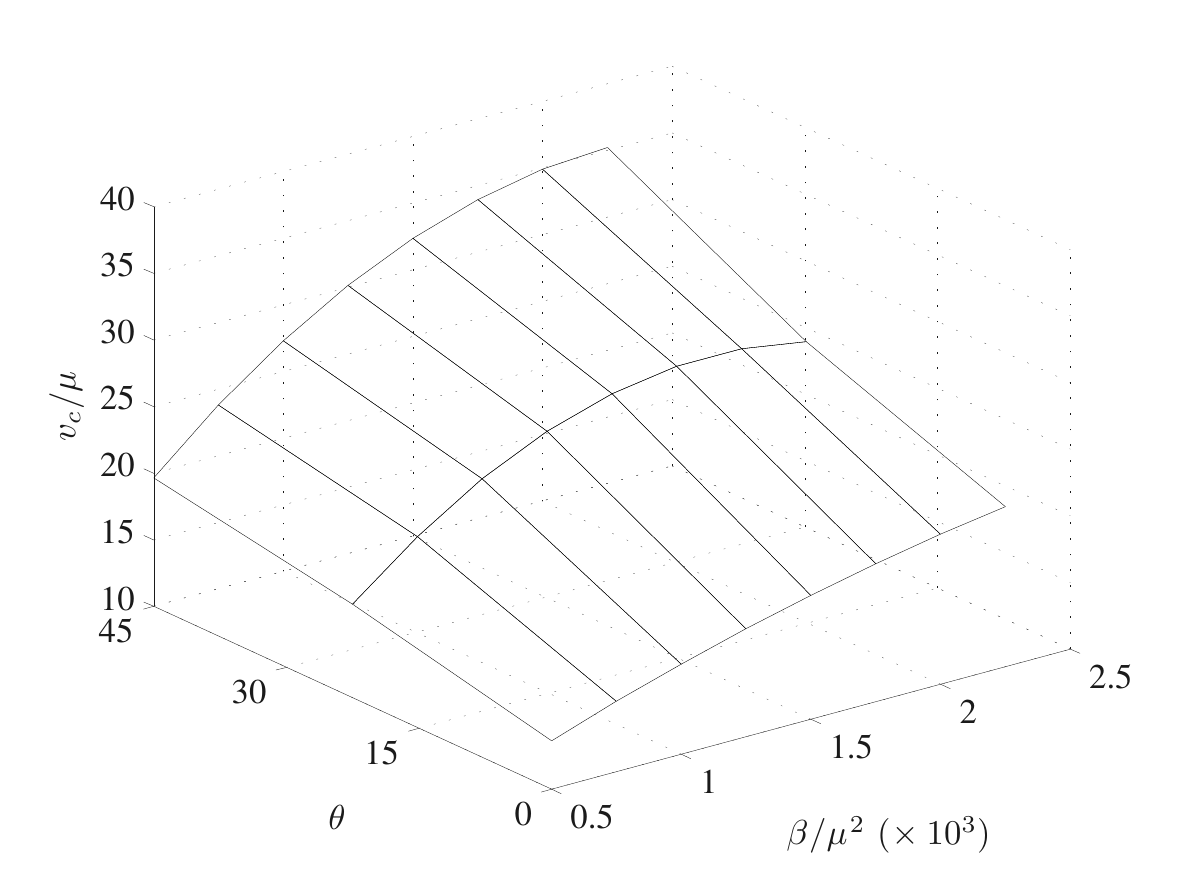}
  \caption{A surface plot of the measured critical flow velocity
    $v_c/\mu= U (M^3 / \rho^2 B)^{1/2}$ versus the in-plane tension
    $\beta / \mu^2 = T ( M^2 /\rho^2 B)$ and its direction
    $\theta$. Cubic polynomial interpolates have been used along the
    $\beta / \mu^2$ axis to re-grid the measured data for mean values of
    $v_c$. This plot pertains to S3 where $\A=0.6$.}
  \label{fig:S3_surface}
\end{figure}

\begin{figure}[p]
  \centering
  \includegraphics[width=0.45\linewidth]{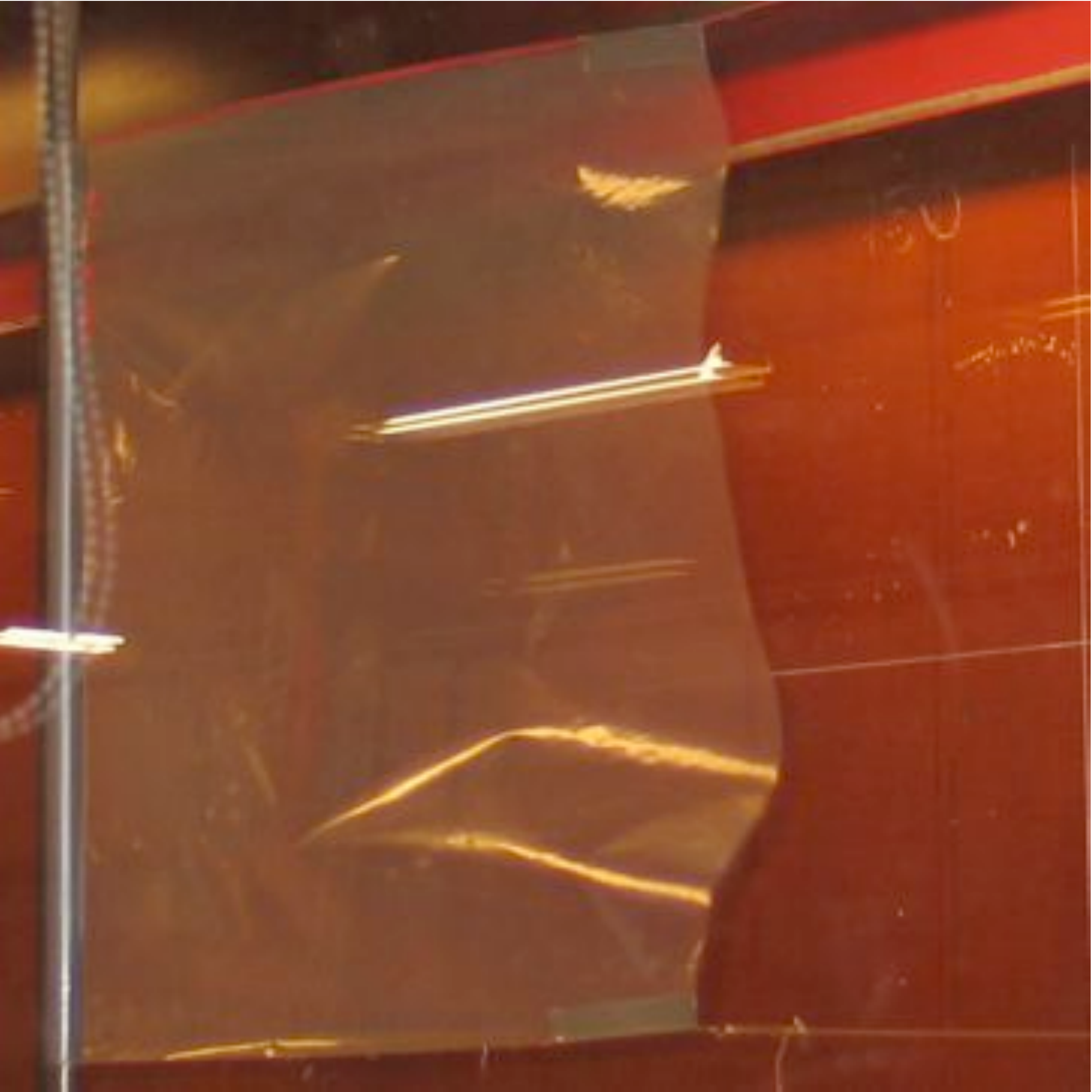}
  \includegraphics[width=0.45\linewidth]{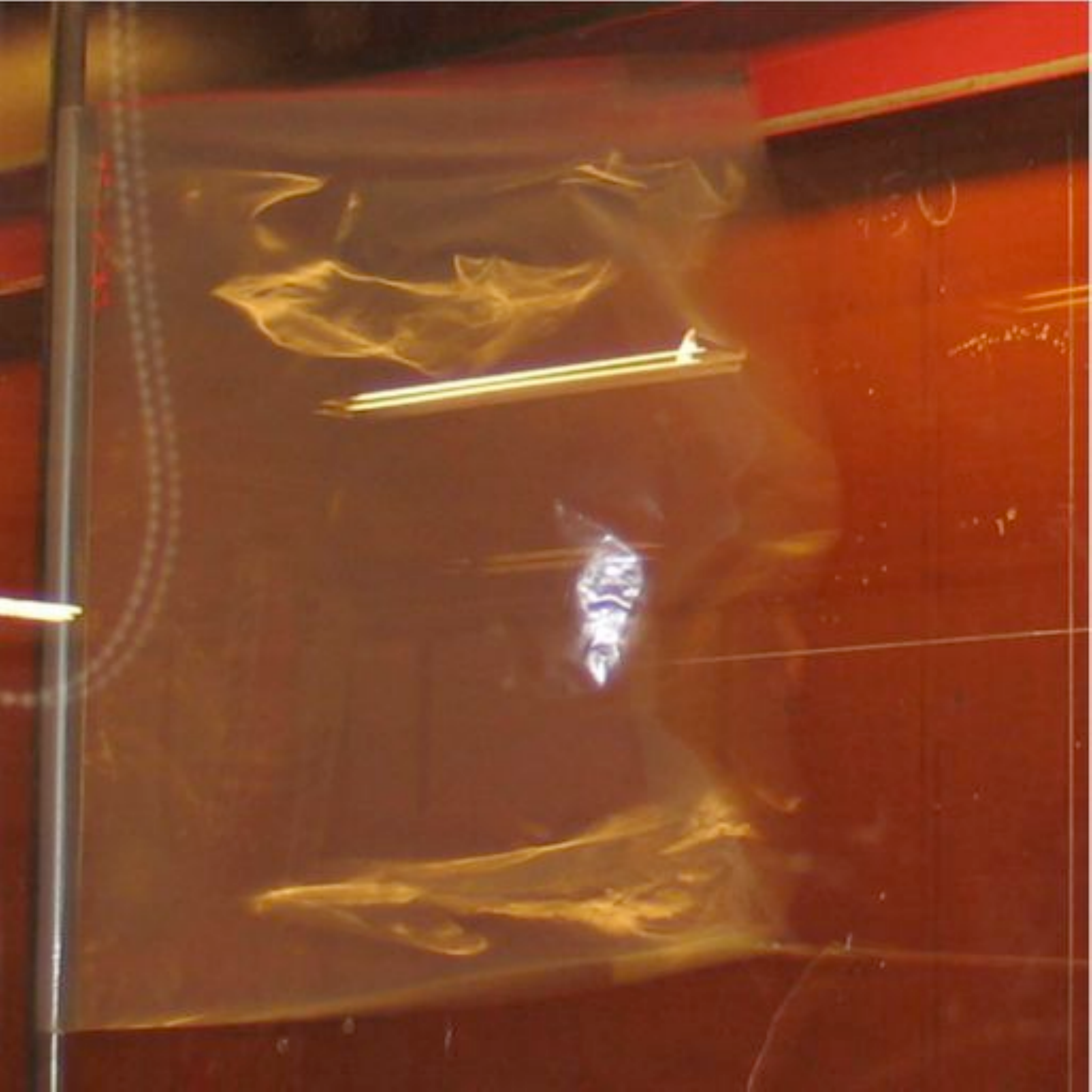}
  \caption{S2 with trailing edge tension $\beta=400$ applied at
    $\theta=0$. A transverse mode is observed across the surface of
    the sheet. Left frame: $v<v_c$; and the Right frame: $v>v_c$.} 
  \label{fig:S2images}
\end{figure}

With the magnitude of the in-plane tension held constant, an increase
in $\theta$ postpones instability. To emphasise this effect, we
present Figure \ref{fig:S3_surface} which shows a surface plot of
$v_c/\mu$ for S3 using a cubic polynomial interpolant over the
measured data presented in Figure \ref{fig:S3_vc}. The distinct trend
of increasing stability with increasing $\theta$, as demonstrated in
the surface plot, is clearly reflected in the measured data for S2 and
S3 (cf. Figures \ref{fig:S2_vc} and \ref{fig:S3_vc}) --- note the
values for $v_c$ at $\beta / \mu^2= 0 $ and $400$ in Figure
\ref{fig:S2_vc}a defy this trend, but, as previously mentioned, the
dynamics of the instability are different here because a transverse
mode shape was excited. Moreover, if one excludes the values of $v_c$
corresponding to $\beta / \mu^2 = 2200$ in Figure \ref{fig:S4_vc}c,
then S4 follows this trend also. For increasing $\theta$, the in-plane
tension comprises an increasing lateral component. Assuming that this
suppresses three-dimensional geometric irregularities, or
perturbations, then one can expect the sheet to adopt a more
streamlined orientation and, consequently, an elevated
stability. However, this trend is unlikely to continue for $\theta >
45\degree$ where increasingly less axial tension would be available to
the sheet.

The effect of $\A$ on the sheet stability is difficult to determine
from the present data set. In most cases, $v_c$ is of the same order of
magnitude for each sheet and no conclusions can be drawn.

\subsection{Fluid dynamic drag of a flag}\label{sec:FD}

\begin{figure}[p]
  \centering
  \includegraphics[scale=1.0]{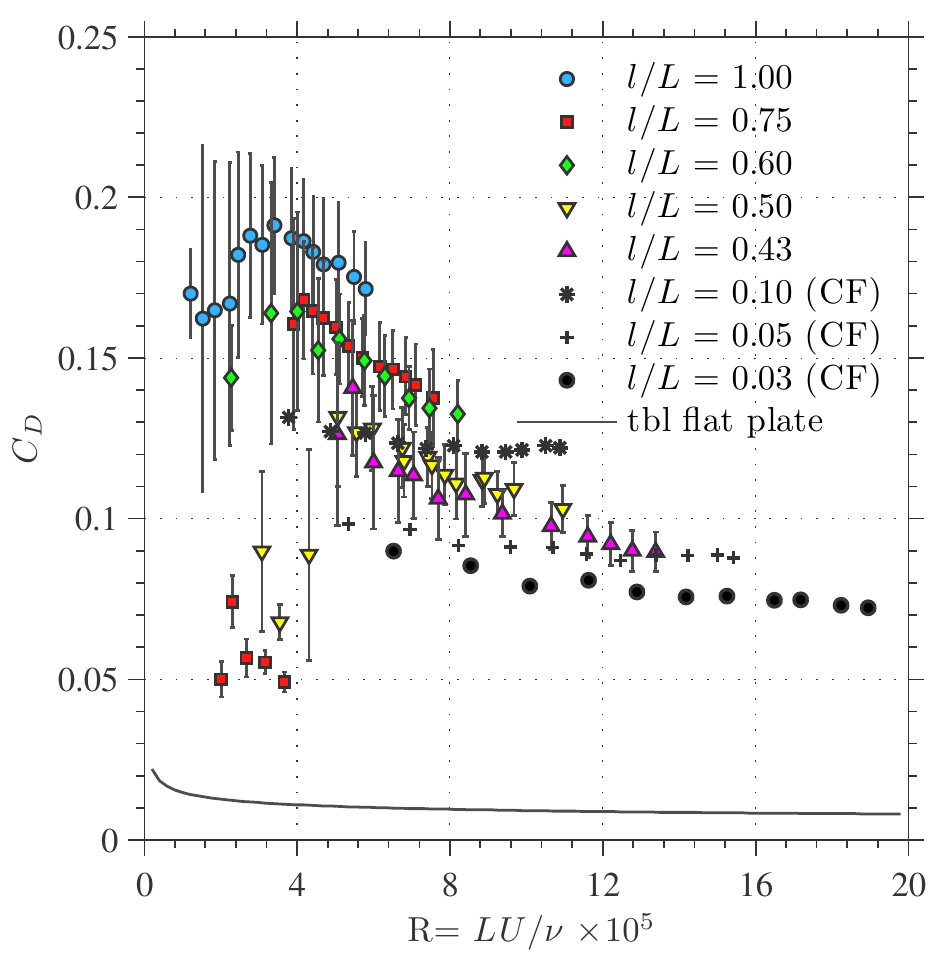}
  \caption{The fluid drag in uniform flow experienced by a horizontal
    flag. The present measurements for a polyethylene sheet are
    plotted for five aspect ratios $\A=l/L$ with error bars
    corresponding to the standard deviation of $C_D$ at each measured
    flow velocity. Published data \cite[]{Carruthers2005}, denoted `CF',
    for a cotton flag provide three additional values of $\A$. The
    solid line represents the expected fluid drag of a flat plate
    experiencing a turbulent boundary layer
    \cite[cf.][eqn. 18.99]{schlt00}.}
  \label{fig:CD_flag}
\end{figure}

We now consider the fluid dynamic drag experienced by flexible sheet
with $\beta=0$ --- in other words, a flag. Measured results are
plotted in Figure \ref{fig:CD_flag} for $C_D= F_D / 0.5 \rho A U^2$
versus the Reynolds number, where the error bars of each data point
correspond to the standard deviation of $C_D$ for each time
window. The results are presented for sheets with aspect ratios $\A =
1,\,0.75,\, 0.6,\, 0.5,\, 0.43$.  In addition to the present
measurements, we compare with published data \cite[]{Carruthers2005},
denoted `CF' hereafter, for a cotton flag comprising aspect ratios $\A
= 0.1,\, 0.05,\, 0.03$.

For comparative purposes, Figure \ref{fig:CD_flag} includes the
expected fluid drag of flat plate experiencing a turbulent boundary
layer in plane parallel flow \cite[cf.][pages~580--586]{schlt00}:
\begin{equation}
  \label{eq:CDt}
  c_D = 2 \left[ \frac{\kappa}{\ln \Ren} \text{G}(\Lambda ;D) \right]^2,
\end{equation}
with
\begin{equation}
  \label{eq:AD}
  \Lambda = \ln \Ren, \qquad D = 2 \ln \kappa + \kappa(C^+ - 3.0),
\end{equation}
where the function $\text{G}(\Lambda; D)$ is determined from the
numerical solution of \citep[eq.~17.60]{schlt00}
\begin{equation}
  \label{eq:G}
  \frac{\Lambda}{\mathrm{G}} + 2 \ln \frac{\Lambda}{\mathrm{G}} - D = \Lambda.
\end{equation}
This friction law is valid for smooth and rough surfaces by judicious
choice of $C^+$ for $\Ren > 10^5$. Given that the sheet material is
polyethylene, it is reasonable to assume a smooth surface and
therefore $C^+=5.0$. The constant $\kappa$ denotes the K\'arm\'an
constant and takes the numerical value 0.41.

As the aspect ratio $\A$ decreases, we observe a decrease in $C_D$
(see Figure \ref{fig:CD_flag}). This trend is also apparent in the
data of CF who studied $\A < 0.1$ --- a range more closely associated
with ribbons and streamers. 

Despite the fact that our $C_D$ measurements are of the same order of
magnitude and follow similar trends as those of CF, a slight vertical
offset in the influence of $\A$ on $C_D$ is evident --- cf. Figure
\ref{fig:CD_flag} and the data sets corresponding to $\A=0.43$ and
CF's $\A=0.1$ for example. Although the surface area of sheets
considered by CF are $\order(10)$ smaller than those examined here,
the mass ratios $\mu$ are quite similar for both studies ($\mu =
5.9-10.2$ for CF and $\mu=6.4-15$ here). However, given that CF
employs cotton flags, the reduced velocity range is of $\order(10^3)$
higher than the present study. This suggests a much greater influence
on the response of the sheet from elastic deformations. As such, this
may account for the slight vertical offset between the two data sets.

At small Reynolds numbers our measurements contain a degree of scatter
which is generally reflected in the magnitude of the error bars. This
uncertainty can be attributed to two factors: the small magnitude of
the drag measured at small $\Ren$; and the sensitivity of $C_D$ to the
sheet's orientation to the flow. For instance, as the flow velocity is
increased the sheet assumes a number of stable non-oscillatory
positions before aligning itself with the flow in a fully extended
position. These changes in orientation affect the frontal and plane
form area exposed to the flow and therefore, by consequence, the fluid
drag. In other words, at low fluid velocities it is reasonable to
expect that the memory or initial conditions of the system can
influence the drag at subsequent fluid velocities before the sheet is
fully extended. Once fully extended, we observe flutter.
 
An example, however, where a fully extended sheet remains stable to
convected waves is demonstrated for S2 where $\A=0.75$. Here, $C_D$
exhibits two distinct regimes centred around a Reynolds number of
$\Ren \approx 4 \times 10^5$. This boundary corresponds to a sharp
rise in $C_D$ from 0.05 to 0.18 which approximates to a scaled flow
velocity of $v/\mu \approx 30$. At this flow velocity the sheet loses
stability to flutter (cf. Figure \ref{fig:S2_vc}a). A similar
phenomenon is demonstrated by S4 with $\A=0.5$, however, the rise in
$C_D$ is far less pronounced. For the remaining sheets, once fully
extended in the flow, stability was lost immediately to flutter.

The measured results indicate that the drag experienced by the flag is
generally of $\order(10)$ greater than that of flat plate of equal
dimensions. Given the sheet's behaviour at small $\Ren$ before reaching
a fully extended orientation in the flow, together with its flutter
response at large $\Ren$, this result not surprising. Within our range
of Reynolds numbers, $\order(10^5-10^6)$, we can expect a turbulent
boundary layer over the surface of the sheet. We notice that, like a
flat plate, the drag coefficient decreases with increasing Reynolds
number. However, the measurements indicate that this feature is
perhaps more pronounced than that experienced by a flat plate. This
would suggest that the boundary layer thickness is larger for a
flexible sheet.

\subsection{Fluid dynamic drag with in-plane tension}\label{sec:tension}

\begin{figure}[p]
  \centering
  \includegraphics{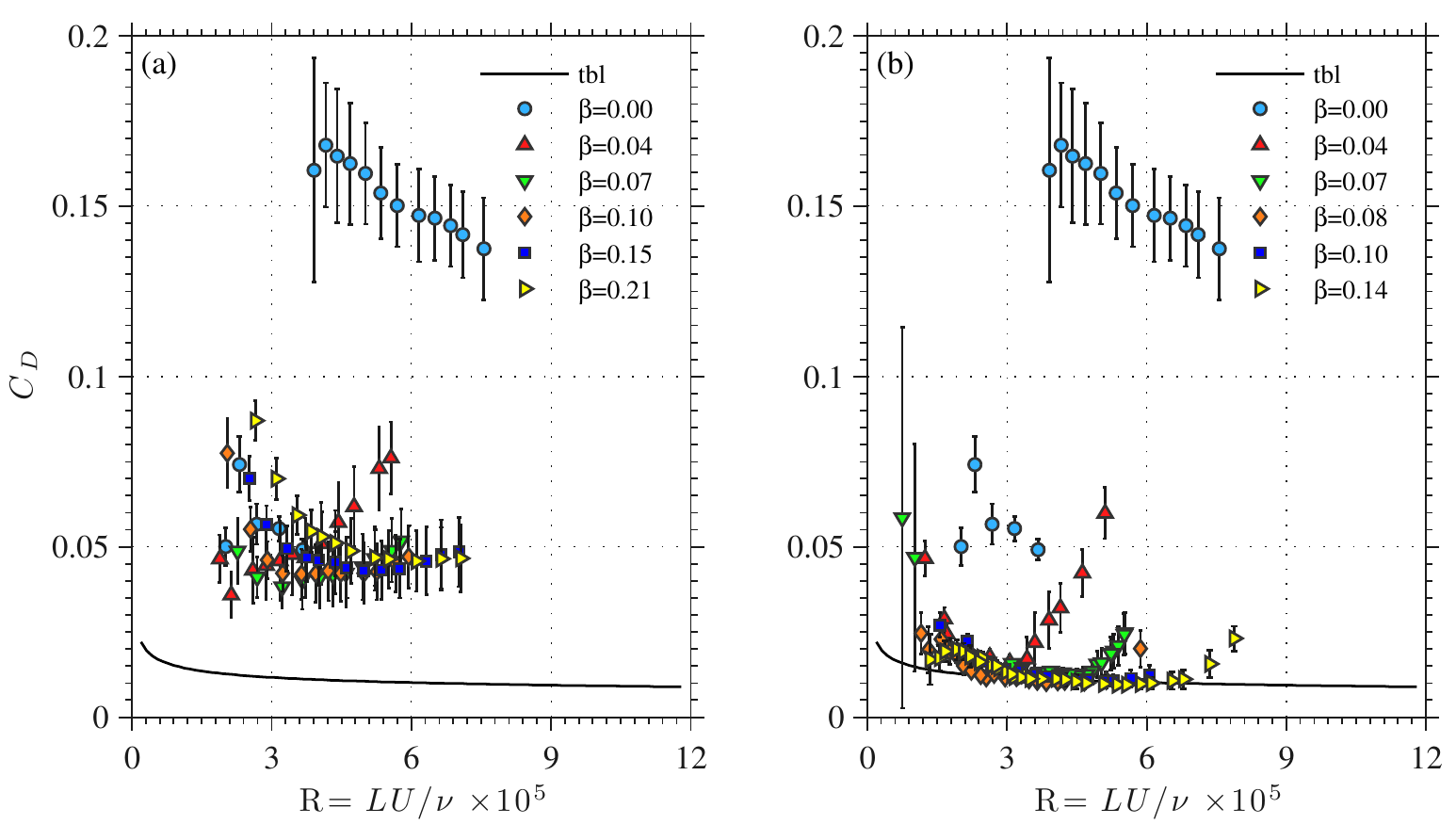}
  \caption{The measured drag of S2 under constant tension
    $\beta=\kappa v^2 \,(\times \, 10^6)$ in uniform flow. The
    subplots correspond to the trailing edge tension applied at
    angles: (a) $\theta=0$; and (b) $\theta=45\degree$. The error bars
    for each data point correspond to the standard deviation of $C_D$
    at each measured flow velocity. The solid line represents the drag
    on a flat plate, of $\A=0.75$, experiencing a turbulent boundary
    layer.}
  \label{fig:CDS2}
\end{figure}

\begin{figure}[p]
  \centering
  \includegraphics{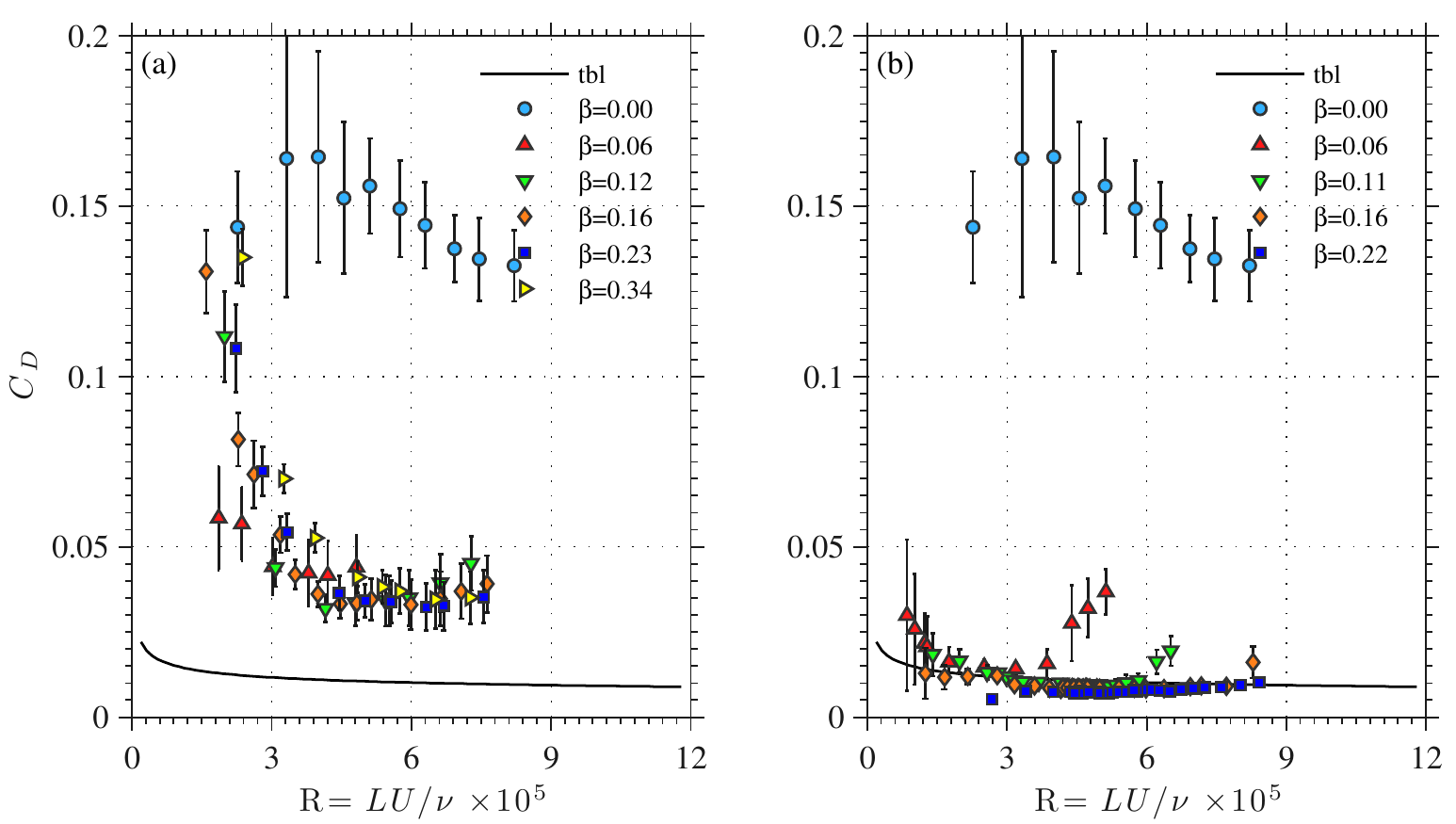}
  \caption{The measured drag of S3 under constant tension
    $\beta=\kappa v^2 \,(\times \, 10^6)$ in uniform flow. The
    subplots correspond to the trailing edge tension applied at
    angles: (a) $\theta=0$; and (b) $\theta=45\degree$. The error bars
    for each data point correspond to the standard deviation of $C_D$
    at each measured flow velocity. The solid line represents the drag
    on a flat plate, of $\A=0.6$, experiencing a turbulent boundary
    layer.}
  \label{fig:CDS3}
\end{figure}

\begin{figure}[p]
  \centering
  \includegraphics{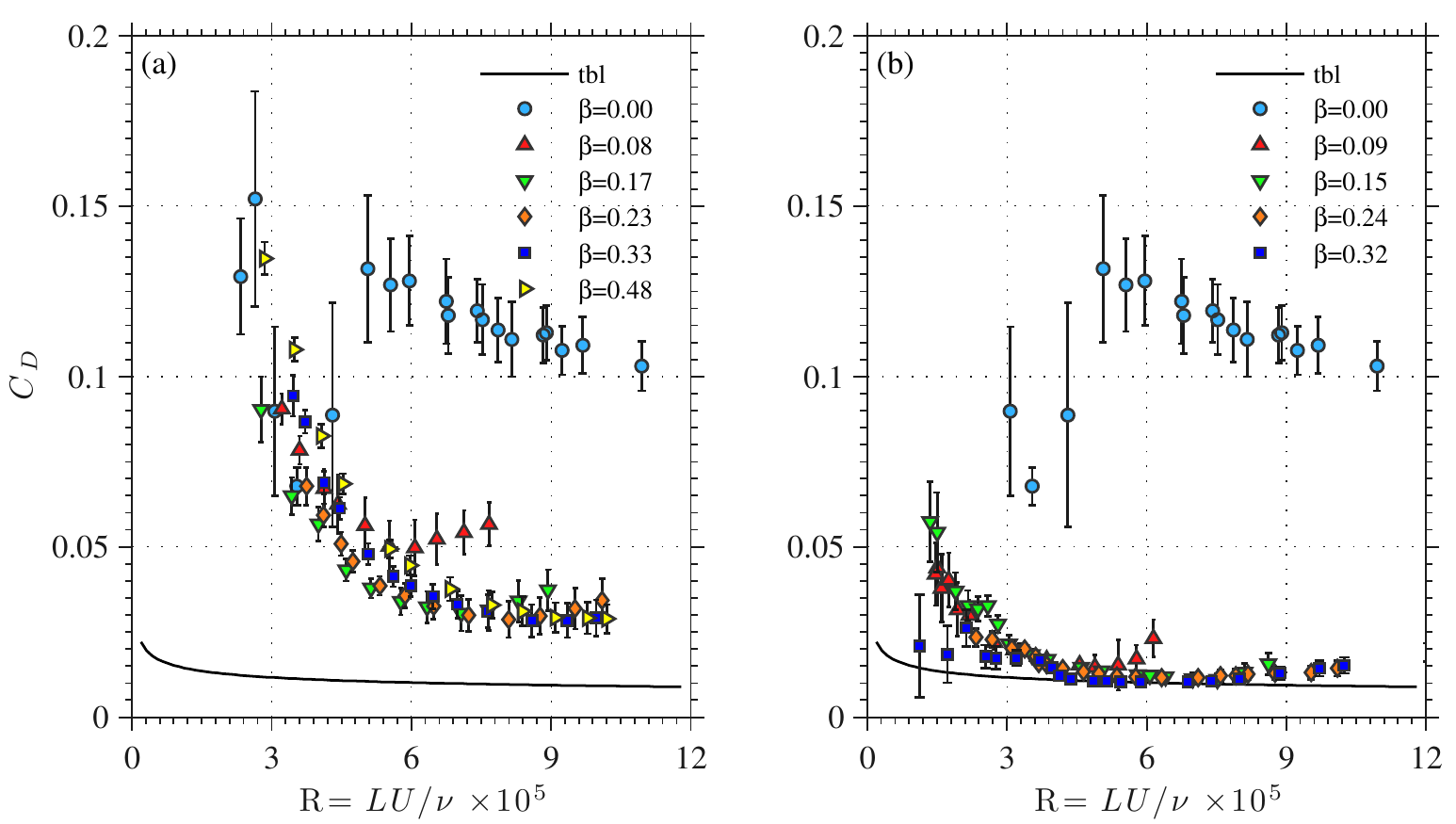}
  \includegraphics{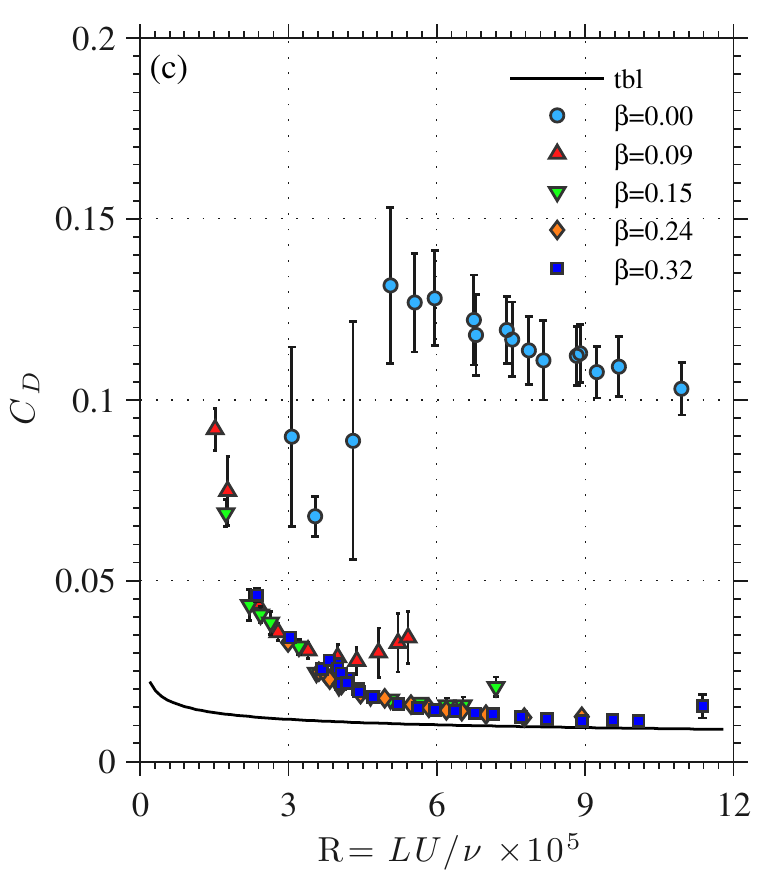}
  \caption{The measured drag of S4 under constant tension
    $\beta=\kappa v^2 \,(\times \, 10^6)$ in uniform flow. The
    subplots correspond to the trailing edge tension applied at
    angles: (a) $\theta=0$; (b) $\theta=22.5\degree$; and (c)
    $\theta=45\degree$. The error bars for each data point correspond
    to the standard deviation of $C_D$ at each measured flow
    velocity. The solid line represents the drag on a flat plate, of
    $\A=0.5$, experiencing a turbulent boundary layer.}
  \label{fig:CDS4}
\end{figure}

We now focus our attention on sheets S2, S3, and S4 and examine the
fluid drag experienced by each in the presence of in-plane tension. We
represent the magnitude of this tension by $\beta = T (L^2 / B)$ which
is applied at an angle $\theta$ to the trailing edge. For S2 and S3,
the drag coefficient is presented for $\theta=0$ and
$\theta=45\degree$ in Figures \ref{fig:CDS2} and \ref{fig:CDS3}
respectively. For S4, however, we consider an additional intermediate
angle of $\theta=22.5\degree$ and these data sets are shown in Figure
\ref{fig:CDS4}. For comparative purposes, the drag experienced by a
flat plate (\ref{eq:CDt}) is plotted on each set of axes.

The measurements demonstrate a significant drag reduction through
a moderate application of in-plane tension at the trailing edge. For
example, S2 with $\beta>0$ experiences a drag reduction by a factor of
approximately $C_D/3$ when compared to that of $\beta=0$ (cf. Figure
\ref{fig:CDS2}). Moreover, such a dramatic reduction depends not only
on the magnitude of the tension, but also on $\theta$. To complicate
matters further, stability also plays a crucial role. Distinctly
different drag characteristics are observed before and after stability
is lost to convected waves.

Before instability $v<v_c$, the magnitude of $C_D$ appears to be
independent of $\beta>0$. Provided sufficient in-plane tension is
applied, the fluid drag of the sheet is reduced regardless of $\beta$
(see Figure \ref{fig:CDS3}b in particular). However, as $\theta$
increases the drag experienced by the sheet is further reduced. This
behaviour is presumably caused by an improved and more streamlined
orientation of the sheet --- this relates to the discussion involving
an increased fluid-elastic stability for increasing $\theta$
(cf. \S~\ref{sec:UC}). This effect is best demonstrated in Figure
\ref{fig:CDS4} for S4. At $\theta=0$, the measured data is perhaps
twice the $C_D$ expected for a flat plate. While for
$\theta=22.5\degree$ and $45\degree$, the drag is reduced to such an
extent that it is given to reasonable approximation by the flat plate
drag law. 

In contrast, for $v \ge v_c$, we observe that the in-plane tension
plays a crucial role in the drag characteristics of the
sheet. Ultimately, the magnitude of the in-plane tension dictates the
critical fluid velocity, and the subsequent rise in the fluid drag
once stability is lost. (cf. Figure \ref{fig:CDS2}b).  This explains
the cascading behaviour of $C_D$, for increasing $\beta$ during
increasing Reynolds numbers, demonstrated in Figures \ref{fig:CDS2}b
and \ref{fig:CDS3}b. Furthermore, these figures also suggest that the
drag behaves in a linear fashion once stability is lost. In
particular, we observe a six fold escalation in $C_D$ past the
critical velocity for a moderate Reynolds number increase of
$10^5$. The convected wave response of the sheet is responsible for
this dramatic drag increase once the critical velocity is
breached. Although not indicated over the range of our data, one would
expect that this linear increase in $C_D$ to be arrested once the
sheet performs chaotic motions and the coherent structures in the wake
of the sheet breakdown \cite[see][Fig. 23 and the related
discussion]{Connell2007} --- this was not investigated here.

\section{Conclusions}\label{sec:conc}

In the present work we have experimentally studied the drag and
stability of a cantilevered flexible sheet in uniform flow
characterised by $\Ren=10^4-10^6$. For the canonical case of a flag,
we observe a decrease in the drag coefficient for decreasing aspect
ratios within the range $0.43<l/L<1$ --- this agrees with previous
studies. In the context of fluid drag and fluid-elastic stability, the
notable feature of this present work has been an investigation into
the effect of in-plane tension applied to the sheet. Our results
indicate that the onset of a flutter-type instability can be delayed
through the application of an in-plane tension. In particular, a
monotonic increase in the critical velocity is observed for increasing
tension. In the presence of in-plane tension, a dramatic reduction in
the drag coefficient of the sheet is observed. The unique drag
characteristics of the sheet are a corollary of its fluid-elastic
stability. Interestingly, two regimes in the drag coefficient are
seen, which appear to be separated by the critical velocity. Before
stability is lost, the drag is well behaved and, provided transverse
modes are suppressed, given to good approximation by the turbulent
boundary layer drag law for a flat plate. However, once the sheet
loses stability, the drag coefficient increases rapidly in a linear
fashion with respect to the onset flow. The point at which this rapid
increase occurs depends on the critical velocity and therefore the
magnitude of the in-plane tension applied to the sheet.
  
\section{Acknowledgements}

The authors gratefully acknowledge the financial support of the
Norwegian Research Council under grant number 169417/530, and
Schlumberger, Norway. Finally, we thank Rune Toennessen, Schlumberger,
for his insightful comments and discussions over the course of this
work.

%
\clearpage

\clearpage

\clearpage

\clearpage

\clearpage

\clearpage


\clearpage
\begin{table}[t]
 \caption{Characteristics of the sheets employed in the test
    matrix where $\A$ denotes the aspect ratio, $A$ the projected
    surface area, $\mu=\rho L/M$ the mass ratio, $v = U (M
    L^2/B)^{1/2}$ the reduced or normalised incident flow velocity,
    $\kappa v^2 = T L^2/B$ the ratio of tension to flexure, $\Ren$ the
    Reynolds number, and $\theta$ represents the angle of application of 
    tension at the trailing edge.}
  \centering
  \begin{tabular}{c|c|c|c|c|c|c|c}
    Sheet & $\mathscr A$ & $A$ (m$^2$) & $\mu$ & $v$ & 
    $\kappa v^2$ $(\times 10^3)$ & $\Ren$ $(\times 10^5)$ &
    $\theta$ \\
    \hline \hline
    S1 &1.0  & 0.563 & 6.410 & 23.0 -- 434 & 19.4 -- 75.0 & 0.315 -- 5.95 & 0 \\
    S2 &0.75 & 0.750 & 8.546 & 37.4 -- 549 & 34.5 -- 133  & 0.513 -- 7.53 & 0, 22.5\degree, 45\degree\\
    S3 &0.60 & 0.938 & 10.68 & 43.2 -- 599 & 54.0 -- 208  & 0.592 -- 8.21 & 0, 22.5\degree, 45\degree\\
    S4 &0.50 & 1.125 & 12.82 & 56.2 -- 744 & 77.7 -- 300  & 0.770 -- 10.2 & 0, 22.5\degree, 45\degree\\
    S5 &0.43 & 1.312 & 14.96 & 63.8 -- 977 & 106  -- 408  & 0.875 -- 13.4 & 0 \\
  \end{tabular}
  \label{tab:parameters}
\end{table}

\end{document}